\documentclass[preprint,aps]{revtex4}
\begin{document}

\draft
\title
{\bf Tunnelling current and emission spectrum of a single electron
transistor under optical pumping}

\author{ David M.-T. Kuo$^1$ and Yia-Chung Chang$^2$}
\address{$^1$Department of Electrical Engineering, National Central
University, Chung-Li, Taiwan, Republic of China}
\address
{$^2$Department of Physics\\
University of Illinois at Urbana-Champaign, Urbana, Illinois
61801}

\date{\today}

\begin{abstract}
Theoretical studies of the tunnelling current and emission
spectrum of a single electron transistor (SET) under optical
pumping are presented. The calculation is performed via Keldysh
Green's function method within the Anderson model with two energy
levels. It is found that holes in the quantum dot (QD) created by
optical pumping lead to new channels for the electron tunnelling
from emitter to collector. As a consequence, an electron can
tunnel through the QD via additional channels, characterized by
the exciton, trion and biexciton states. It is found that the
tunnelling current as a function of the gate voltage displays a
series of sharp peaks and the spacing between these peaks can be
used to determine the exciton binding energy as well as the
electron-electron Coulomb repulsion energy. In addition, we show
that the single-photon emission associated with the electron-hole
recombination in the exciton complexes formed in the QD can be
controlled both electrically and optically.
\end{abstract}

\maketitle


\section{Introduction}
    The tunnelling current characteristics of a single electron
transistor (SET) have been extensively studied and interesting
physical phenomena, such as Coulomb blockade and Kondo effect have
been found.$^{1-7}$ The main structure of an SET consists of three
electrodes (source, drain and gate) and the active region consists
of one or more quantum dots (QDs). Although the physics of Kondo
effect is more profound than that of Coulomb blockade, the main
device application of SET is based on the Coulomb blockade effect.
The tunnelling current of an SET is sensitive to the size of the
QD. The charging energy, $\Delta U$, is typically larger than the
energy level spacing, $\Delta E $, caused by the quantum
confinement in large QDs. Consequently, the tunnelling current
displays a periodic Coulomb oscillation with respect to the gate
voltage. This feature of periodic Coulomb oscillation is a
consequence of homogeneity of $\Delta U$ and negligible $\Delta
E$. However, such a Coulomb oscillation was observed only at very
low temperature due to the small charging energy. Note that the
charging energy must be greater than the thermal energy and
tunnelling rate multiplied by $\hbar$ in order for an SET to
operate properly. For application in high temperature SETs,
nanoscale QDs with large charging energy are desired.

To date two major kinds of nanoscale QDs have been used to
construct SETs: the Si/Ge QDs and the InAs/GaAs self-assembled
quantum dots (SAQDs). Silicon (Si) or germanium (Ge) QDs can be
miniaturized to reach nanometer scale with advanced fabrication
technology. Nanoscale Si/Ge SETs typically display non-periodic
Coulomb oscillations at room temperatures,$^{8-11}$ as a result of
the large charging energy ($\Delta U$) and the unequally spaced
values of $\Delta E$, which are comparable to $\Delta U$. Even
though Si/Ge SETs can operate at room temperatures, it is
difficult to gian full understanding of their tunnelling current
characteristics because of the multi-valleyed nature of Si/Ge and
the unknown Si/Ge-SiO$_2$ interface properties. Furthermore, the
electron-hole recombination rate in Si/Ge QDs is not large enough
for useful application in optoelectronics, since Si and Ge are
indirect-band-gap semiconductors.$^{12}$

     The InAs/Gas SAQDs, on the other hand, are better understood,
and because InAs is a direct-band-gap semiconductors, SETs
constructed from  InAs/Gas SAQDs can have useful application in
optoelectronic devices such as single-photon generators and
single-photon detectors. Recently, InAs/GaAs self-assembled
quantum dot (SAQD) embedded in a p-n junction has been employed to
generate single-photon emission for application in quantum
cryptography.$^{13-15}$  Another way to achieve single-photon
generation is to use isolated SAQD with optical pumping. Such
studies have received a great deal of attention
recently.$^{16-21}$ The spontaneous emission spectrum of a QD
typically exhibits coexisting sharp emission peaks, which have
been attributed to the electron-hole recombination in the exciton,
trion and biexciton formed in the QD. The antibunching feature of
the emitted single photon in this kind of device has been
demonstrated.$^{13,14}$ In addition, gate electrodes have been
used in some experiments to manipulate the emission
spectrum.$^{22-24}$.

   Many theoretical efforts
based on the Anderson model and Keldysh Green's function method
have been devoted to the studies of transport properties of SETs,
including the dc and ac tunnelling current.$^{5-7}$ The studies on
the ac tunnelling current of SET mainly focus on the photon-side
band phenomena caused by the microwave pumping$^{5,7}$. Only a
handful theoretical studies considered the optical properties of
SETs. Recently, we have studied the intraband transitions of SETs
made of InAs/GaAs SAQDs for application as an infrared
photodetector with operating wavelength near $10 \mu m$.$^{25,26}$
On the other hand, the interband transitions in a QD have been
adopted frequently to realize single-photon generation.
Theoretical studies of the electronically driven single-photon
generator (SPG) have been reported,$^{27,28}$ whereas theoretical
studies on optically pumped SETs (including the coupling with
leads) are still lacking. It is the purpose of this paper to
provide a theoretical analysis for such a system. Furthermore, the
measurement of exciton binding energy in a QD remains illusive,
since in a typical photoluminescence measurement or the emission
spectrum of a single QD, one can only observe the exciton
recombination, but not the free electron-hole recombination. The
exciton binding energy is an important physical quantity as it
provides information about the effective dielectric constant and
the charge distribution of electron and hole in the QD. In our
current theoretical study, we will show that via the measurement
of the tunnelling current of an SET made of a nanoscale QD under
optical pumping one can determine the exciton binding energy of a
QD unambiguously.

    The schematic diagram for the SET system considered is shown in
Fig. 1. Here a nanoscale QD is embedded in a n-i-n junction. This
is different from the electrically driven SPG, where a QD is
embedded in a p-n junction. Therefore, the physical process
considered here is also quite diffrent. Using optical pumping to
create electron-hole pairs in the SET, one can not only
electrically manipulate the single-photon emission spectrum
arising from exciton complexes in the QD, but also optically
control the tunnelling current of the SET. In this paper, we will
consider both the hole-assisted tunnelling current due to optical
pumping and the electrode-controlled spontaneous emission
spectrum.

    Our calculation is based on the Keldysh Green's function
approach$^{29}$ within the Anderson model for a two-level system.
A effective-mass model, which takes into account the anisotropy
and inhomogeneity, is used to estimate the inter-particle Coulomb
energies. We find that the optical excitation creates holes in the
QD, which provide new channels (via the electron-hole interaction)
for the electron to tunnel from the emitter to the collector. As a
consequence, an electron can tunnel through the QD via four
additional channels, characterized by the exciton, positive trion,
negative trion, and biexciton states. Each addition channel can
generate a new plateau (or oscillatory peaks) in the tunnelling
current characteristics in addition to the typical plateau (or
peaks) caused by the Coulomb blockade effect. This gives rise to a
rich tunnelling current characteristics and it can be used to
determine the exciton binding energy.

This paper is organized as the following. In section II, we derive
the tunnelling current of SET under optical pumping. In section
III, we calculate the polarization needed for describing the
spontaneous emission spectrum. In section IV, we present
calculations of the inter-particle Coulomb interactions within a
simple but realistic effective-mass model. In section V, we
discuss the results of our numerical calculations on the
tunnelling current and spontaneous emission spectrum and
demonstrate that the exciton binding energy can be extracted via
the measurement of tunnelling current. Finally, a summary and
concluding remarks are presented in section VI.

\section{Tunnelling current}
The system of interest is shown in Fig. 1, which consists of a
single InAs/GaAs self-assembled quantum dot (SAQD) sandwiched
between two n-doped GaAs leads. Electrons are allowed to tunnel
from the left lead (emitter) to the right lead (collector) under
the influence of an optical field.  The Hamiltonian for the system
is given by

\begin{eqnarray}
H&=& \sum_{\sigma,i=e,h}E_i d^{\dagger}_{i,\sigma} d_{i,\sigma}+
\sum_{{\bf k},\sigma;\ell=L,R} \epsilon_{\bf k} c^{\dagger}_{{\bf
k},\sigma;\ell}c_{{\bf k},\sigma;\ell} \nonumber \\ & + &
\sum_{\ell=L,R;k,\sigma}(V_{{\bf k},\sigma;\ell} c^{\dagger}_{{\bf
k},\sigma;\ell} d_{e,\sigma}+ V^{\dagger}_{{\bf k},\sigma;\ell}
d^{\dagger}_{e,\sigma} c_{{\bf k},\sigma;\ell}) \nonumber\\ & +&
\sum_{{\bf k},\sigma} (\lambda_0 e^{i\omega_0 t} b_{e,{\bf
k},\sigma} b_{h,{\bf k},-\sigma}+ \lambda_0^{\dagger} e^{-\omega_0
t} b^{\dagger}_{h,{\bf k}, -\sigma} b^{\dagger}_{e,{\bf k},
\sigma})\nonumber\\ & +& \sum_{i,{\bf k}, {\bf q},\sigma} (
g_{i,{\bf q}} A b^{\dagger}_{i,{\bf k},\sigma} d_{i,\sigma}+
g^{\dagger}_{i,{\bf q}} A^{\dagger}
d^{\dagger}_{i,\sigma} b_{i,{\bf k},\sigma} )  \nonumber\\
&+&\sum_{{\bf Q}}(\lambda e^{i \omega t} a^{\dagger} d_{e,\sigma}
d_{h,-\sigma}+ \lambda^{\dagger} e^{-i \omega t} a
d^{\dagger}_{h,-\sigma} d^{\dagger}_{e,\sigma}),
\end{eqnarray}
where the first term describes electrons in the QD. We assume that
the quantum confinement effect is strong for the small InAs QD
considered here. Therefore, the energy spacings between the ground
state and the first excited state for electrons and holes, $\Delta
E_{e}$ and $\Delta E_{h}$, are much larger than the thermal
energy, $k_B T$, where $k_B$ and $T $ denote the Boltzmann
constant and temperature. Only the ground state levels for
electrons and holes, $E_e$ and $E_h$, are considered in the first
term. The second term describes the kinetic energies of free
carriers in the electrodes. Note that in the current setup, the
gate electrode does not provide any carriers, but merely controls
the energy levels of the QD. The third term describes the coupling
between the QD and the leads. Note that only the conduction band
of QD is coupled with the electrodes. The fourth term describes
the interband optical pumping with a frequency $\omega_0$, which
is in resonance with the transition energy for an electron-hole
pair in the wetting layer. We treat the electromagnetic field as a
semiclassical field. The fifth term describes the optical phonon
assisted process for electrons in the wetting layer to relax into
the InAs QD. The last term describe the coupling of the QD with
the electromagnetic field of frequency $\omega$. $\lambda\equiv
-\mu_r {\cal E}$ is the Rabi frequency, where $\mu_r=<f|\bf{r}|i>$
is the matrix element for the optical transition and ${\cal E}$ is
electric field per photon. $a^{\dagger} (A^{\dagger})$ and $a$
($A$) denote the creation and annihilation operators of a photon
(phonon), respectively.

    Due to the large strain-induced splitting between the heavy-hole
and light-hole bands for typical QDs, we only have to consider the
heavy hole band (with $J_z=\pm 3/2$) and ignore its coupling with
light-hole band caused by the QD potential. Thus, we can treat the
heavy hole as a spin-1/2 particle with $\sigma= \downarrow,
\uparrow$ representing $J_z = \pm 3/2$. This treatment is
convenient for algebraic manipulations in the calculation of
Green's functions. Because the effect of inter-particle Coulomb
interactions is significant in small semiconductor QDs, we take
into account the electron Coulomb interactions and electron-hole
Coulomb interactions by adding the following terms
\begin{equation}
H_I = \sum _{i,\sigma} U_{i,i} d^{\dagger}_{i,\sigma} d_{i,\sigma}
d^{\dagger}_{i,-\sigma}d_{i,-\sigma}+ \sum_{i \neq j;\sigma,
\sigma'}U_{i,j} d^{\dagger}_{i,\sigma} d_{i,\sigma}
d^{\dagger}_{j,-\sigma}d_{j,-\sigma},
\end{equation}
where $d^{\dagger}_{i,\sigma}$ and $d_{i,\sigma}$ are creation and
annihilation operators for particles with spin $\sigma$ in the
$i$th energy level of the QD.  Once the Hamiltonian is
constructed, the tunnelling current of SET can be calculated via
the Keldysh Green's function method.$^{27}$ We obtain the
tunnelling current through a single dot (see appendix A)
\begin{equation}
J=\frac{-2e}{\hbar}\int \frac{d\epsilon}{2\pi}[f_L(\epsilon-\mu_L)
-f_R(\epsilon-\mu_R)]\frac{\Gamma_L(\epsilon) \Gamma_R(\epsilon)}
{\Gamma_L(\epsilon)+\Gamma_R(\epsilon)}ImG^r_{e,\sigma}(\epsilon),
\end{equation}
where $f_L(\epsilon)$ and $f_R(\epsilon)$ are the Fermi
distribution function for the source and drain electrodes,
respectively. The chemical potential difference between these two
electrodes is related to the applied bias via $\mu_L-\mu_R=eV_a$.
$\Gamma_L(\epsilon)$ and $\Gamma_R(\epsilon)$ denote the
tunnelling rates from the QD to the left (source) and right
(drain) electrodes, respectively.
For simplicity, these tunnelling rates will be assumed energy and
bias-independent. Therefore, the calculation of tunnelling current
is entirely determined by the spectral function
$A=ImG^r_{e,\sigma}(\epsilon)$, which is the imaginary part of the
retarded Green's function $G^r_{e,\sigma}(\epsilon)$. Eq.~(3) is
obtained provided that the condition $\Gamma_{L(R)}>\gg
\gamma_{e,c}$ and $\Gamma_{L(R)} \gg R_{eh} $ are satisfied, where
$\gamma_{e,c}$ is the captured rate of electrons. Because of the
phonon bottleneck effect, we expect the rate for the electron to
relax from the wetting layer to the QD is small. Thus, we have
ignored the terms involving phonons in the calculation of the
tunnelling current. Besides, photon current has also been
neglected, since the electron-hole recombination rate, $R_{eh}$ is
much smaller than the tunnelling rates.

The retarded Green's function, $G^{r}_{e,\sigma}(\epsilon)$, can
be obtained by the equation of motion of
$G^r_{e,\sigma}(t)=-i\theta(t)\langle
\{d_{e,\sigma}(t),d^{\dagger}_{e,\sigma}(0)\}\rangle$, where
$\theta(t)$ is a step function. The curly brackets represent the
anti-commutator and the bracket  $\langle ...\rangle  $ denotes
the thermal average. The Fourier transform of $G^r_{e,\sigma}(t)$
is given by
\begin{equation}
G^{r}_{e,\sigma}(\epsilon)=\int_{-\infty}^{\infty}dt
G^{r}_{e,\sigma}(t) e^{i(\epsilon+i\eta)t}
\end{equation}
with $\eta$ being a positive infinitesimal number.
$G^{r}_{e,\sigma}$ is the full Green's function, which includes
all type of interactions. Given the conditions $\Gamma_{L(R)} \gg
\gamma_{e,c} $ and $\Gamma_{L(R)} \gg R_{eh} $, we can drop the
fifth and the last term of the Hamiltonian in the calculation of
$G^{r}_{e,\sigma}(\epsilon)$. To solve
$G^{r}_{e,\sigma}(\epsilon)$, we consider only the lowest order
coupling between the electrodes and the QD. The equation of motion
for $G^{r}_{e,\sigma}(t)$ leads to

\begin{small}
\begin{equation}
(\epsilon-E_e+i\frac{\Gamma_{e}}{2})G^r_{e,\sigma}(\epsilon)=1+U_e
G^r_{ee}(\epsilon) -U_{eh}(
G^r_{eh,1}(\epsilon)+G^r_{eh2}(\epsilon)),
\end{equation}
\end{small}
where the two particle Green's functions, $G^r_{ee}(\epsilon)$,
$G^r_{eh1}(\epsilon)$ and $G^r_{eh,2}(\epsilon)$ arise from the
particle correlation and they satisfy
\begin{small}
\begin{equation}
(\epsilon-(E_e+U_e)+i\frac{\Gamma_e}{2})G^r_{ee}(\epsilon)
=N_{e,-\sigma}-U_{eh}(G^r_{ehe1}(\epsilon)+ G^r_{ehe2}(\epsilon)),
\end{equation}
\begin{equation}
(\epsilon-(E_e-U_{eh})+i\frac{\Gamma_e}{2})
G^{r}_{eh1}(\epsilon)=n_{h,\sigma} +U_{e}
G^r_{ehe1}(\epsilon)-U_{eh} G^r_{ehh}(\epsilon),
\end{equation}
\end{small}
and
\begin{small}
\begin{equation}
(\epsilon-(E_e-U_{eh})+i\frac{\Gamma_e}{2}) G^r_{eh2}(\epsilon)=
n_{h,-\sigma} +U_{e}
G^r_{ehe2}(\epsilon)-U_{eh}G^r_{ehh}(\epsilon).
\end{equation}
\end{small}
$N_{e,-\sigma}$, $n_{h,\sigma}$ and $n_{h,-\sigma}$ in
Eqs.~(6)-(8) denote the steady-state electron and hole occupation
numbers. The two-particle Green's functions are coupled with the
three-particle Green's functions defined as
$G^r_{ehe1}(\epsilon)=\langle{N_{e,-\sigma} n_{h,\sigma}
d_{e,\sigma},d^{\dagger}_{e,\sigma}}\rangle$,
$G^r_{ehe2}(\epsilon)=\langle{N_{e,-\sigma} n_{h,-\sigma}
d_{e,\sigma},d^{\dagger}_{e,\sigma}}\rangle$ and
$G^r_{ehh}(\epsilon)=\langle{n_{h,-\sigma}n_{h,\sigma}
d_{e,\sigma},d^{\dagger}_{e,\sigma}}\rangle$. The equations of
motion of the three-particle Green's functions will lead to
coupling with the four-particle Green's functions (two electrons
and two holes), where the hierarchy terminates. Thus, these
three-particle Green's functions can be expressed in the following
closed form
\begin{eqnarray}
G^r_{ehe1}(\epsilon) =
N_{e,-\sigma}n_{h,\sigma}(\frac{1-n_{h,-\sigma}}
{\epsilon-(E_e+U_{e}-U_{eh})+i\frac{\Gamma_e}{2}}\nonumber \\
+\frac{n_{h,-\sigma}}
{\epsilon-(E_e+U_{e}-2U_{eh})+i\frac{\Gamma_e}{2}}),
\end{eqnarray}

\begin{eqnarray}
G^r_{ehe2}(\epsilon)=N_{e,-\sigma}n_{h,-\sigma}(\frac{1-n_{h,\sigma}}
{\epsilon-(E_e+U_{e}-U_{eh})+i\frac{\Gamma_e}{2}} \nonumber \\
+ \frac{n_{h,\sigma}}
{\epsilon-(E_e+U_{e}-2U_{eh})+i\frac{\Gamma_e}{2}}),
\end{eqnarray}
and
\begin{eqnarray}
G^r_{ehh}(\epsilon)= n_{h,-\sigma} n_{h,\sigma}
(\frac{1-N_{e,-\sigma}}{\epsilon-(E_e-2U_{eh})+i\frac{\Gamma_e}{2}}\nonumber \\
+\frac{N_{e,-\sigma}}{\epsilon-(E_e-2U_{eh}+U_{e})+i\frac{\Gamma_e}{2}}).
\end{eqnarray}
Eqs.~(9) and (10) describe  the mixed amplitudes for the
propagation of an electron either in the presence of another
electron (with opposite spin) plus one hole, or another electron
plus two holes. Eq.~(11) describes the mixed amplitudes for the
propagation of an electron either in the presence of two holes, or
two holes plus another electron. Substituting Eqs.~(9)-(11) into
Eqs.~(7) and (8), we obtain, after some algebras, the retarded
Green's function of Eq.~(3)
\begin{small}
\begin{eqnarray}
&  & G^r_{e,\sigma}(\epsilon)=
(1-N_{e,-\sigma})\{\frac{1-(n_{h,\sigma}+n_{h,-\sigma})+n_{h,\sigma}n_{h,-\sigma}}
 {\epsilon - E_{e} +i \frac{\Gamma_{e}}{2} }\nonumber \\ \nonumber &
+&\frac{n_{h,\sigma}+n_{h,-\sigma}-2
n_{h,\sigma}n_{h,-\sigma}}{\epsilon - E_{e} + U_{eh} +
i\frac{\Gamma_e}{2}} +\frac{n_{h,\sigma}n_{h,-\sigma} }{\epsilon -
E_{e} + 2U_{eh}+ i\frac{\Gamma_{e}}{2}}\}\\ \nonumber& + &
N_{e,-\sigma}\{\frac{1-(n_{h,\sigma}+n_{h,-\sigma})+n_{h,\sigma}n_{h,-\sigma}}
{\epsilon - E_{e}- U_{e} + i\frac{\Gamma_{e}}{2}} \nonumber \\&+&
\frac{n_{h,\sigma}+n_{h,-\sigma}-2
n_{h,\sigma}n_{h,-\sigma}}{\epsilon-
E_{e}- U_{e}+U_{eh} + i\frac{\Gamma_{e}}{2}} \nonumber \\
 &+ & \frac{n_{h,\sigma}n_{h,-\sigma}}{\epsilon - E_{e}-
U_{e}+ 2U_{eh}+ i\frac{\Gamma_{e}}{2}} \}.
\end{eqnarray}
\end{small}
Here $G^r_{e,\sigma}(\epsilon) $ contains an admixture of six
possible configurations in which a given electron can propagate.
These configurations are: empty state, one-hole state, two-hole
state, one-electron state, one-electron plus one-hole state, and
one-electron plus two-hole state. In Eq.~(12), $\Gamma_e$ is the
electron tunnelling rate $\Gamma_e \equiv \Gamma_L+\Gamma_R$. The
particle correlation has not been included in the tunneling rates
of Eq. (12). Such approximation has been known to be adequate for
describing the Coulomb blockade effect, but not the Kondo effect.
The electron occupation number of the QD can be solved
self-consistently via the relation

\begin{small}
\begin{equation}
N_{e,\sigma} = -\int \frac{d\epsilon}{\pi} \frac{\Gamma_L
f_L(\epsilon)+\Gamma_R f_R(\epsilon)}{\Gamma_L+\Gamma_R}
ImG^r_{e,\sigma}(\epsilon).
\end{equation}
\end{small}
$N_{e,\sigma}$ is limited to the region $0 \le N_{e,\sigma} \le 1
$. Eq.~(13) indicates that the electron occupation numbers of the
QD, $N_{e,-\sigma}$ and $N_{e,\sigma}$, are primarily determined
by the tunnelling process. To obtain the electron and hole
occupation numbers ($n_{e,-\sigma}=n_{e,\sigma}$ and
$n_{h,-\sigma}=n_{h,\sigma}$) arising from the optical pumping, we
solve the rate equations (see appendix B) and obtain
\begin{equation}
n_e=n_{e,-\sigma}=n_{e,\sigma}=\frac{\gamma_{e,c} N_{e,{\bf
k}}(1-N_e)} {\gamma_{e,c} N_{e,{\bf k}} + R_{eh} n_{h}+ \Gamma_e}
\end{equation}
and
\begin{equation}
n_h=n_{h,-\sigma}=n_{h,\sigma}=\frac{\gamma_{h,c} N_{h,{\bf k}}}
{\gamma_{h,c} N_{h,{\bf k}} + R_{eh}(n_e + N_e)+ \Gamma_{h,s} },
\end{equation}
where $\gamma_{e(h),c}$ and $N_{e(h),{\bf k}}$ denote the captured
rate for electrons (holes) from the wetting layer to the QD and
the occupation number of electrons (holes) in the wetting layer.
See Eqs. (37) and (38), $N_{e(h),{\bf k}}$ is a linear function of
$ p_{exc}$ for low pumping intensity and small $\gamma_{e(h),c}$.
$\Gamma_{h,s}$ denotes the non-radiative recombination rate for
holes in the QD. Because $\gamma_{e,c}N_{e,{\bf k}}/\Gamma_e \ll
1$, $n_e$ has been neglected in Eq. (12).

\section{Spontaneous emission spectrum}
When electrons and holes appear in the QD, their recombination
leads to single-photon emission. The optical polarization of the
emitted photon depends on the spin polarization of the electrons.
An electron with spin +1/2 (-1/2) can recombine with a heavy hole
of angular momentum +3/2 (-3/2) to emit a circularly ($\sigma^{+}
or \sigma^{-}$) polarized photon. In some experiments, up to four
coexisting peaks (associated with exciton, negative trion,
positive trion and biexciton) have been observed in the emission
spectrum.$^{16-18}$ The emission process is described by the last
term in the Hamiltonian given in Eq.~(1). The bias-dependent
emission spectrum can be obtained by finding the nondiagonal
lesser Green's function at nonequal time.$^{7}$ To simplify the
problem, we assume that the applied bias is large enough to create
the biexciton state (two electrons and two holes) in the QD.

With the above assumption, polarization $P(\omega)=
G^{<}_{eh}(\omega)$ can be calculated by the equal-time Keldysh
Green function, $G^{<}_{eh}(t,t)=-i\langle
a^{\dagger}(t)d_{h,-\sigma}(t)d_{e,\sigma}(t)\rangle $ (or density
matrix method). Prior to the calculation of the polarization
$P(\omega)$, a unitary transformation has been used to remove the
phase of the optical transition term. The renormalized energy
levels of the electron and hole become
$\epsilon_e=E_e-\frac{\omega}{2}$ and
$\epsilon_h=E_h-\frac{\omega}{2}$.  Furthermore, the hopping terms
between the leads and dot, $V_{j=L,R;{\bf k}}(t) =V_{j=L,R;{\bf k}
} exp^{-\frac{(\pm i \omega t)}{2}}$ become time dependent, in
which the energy and time dependence of the coupling are
factorized. This factorization leads to time-independent
tunnelling rates and it simplifies the calculation. Solving the
equation of motion for $G^{<}_{eh}(t,t)=-i\langle
a^{\dagger}(t)d_{h,-\sigma}(t)d_{e,\sigma}(t)\rangle $ and
considering the steady-state condition, we obtain the lesser
Green's function

\begin{equation}
G^{<}_{eh}(\omega)= \frac{-\lambda {\cal F}-(U_e-U_{eh}){\cal
P}_1(\omega)-(U_h-U_{eh}){\cal P}_2(\omega)}
{\epsilon_e+\epsilon_h-U_{eh}+i\frac{\Gamma}{2}},
\end{equation}
where
\begin{equation}
{\cal F} =\langle \Phi|a^{\dagger}a|\Phi\rangle (1-N_{e,\sigma}-
n_{h,-\sigma})-N_{e,\sigma} n_{h,-\sigma},
\end{equation}
$\Gamma\equiv \Gamma_e+\Gamma_h$, which leads to a broadening of
the spectrum, and $|\Phi\rangle$ denotes a photon state. ${\cal
P}_1(\omega)$ and ${\cal P}_2(\omega)$ are given by

\begin{eqnarray}
{\cal P}_1(\omega)=-\lambda {\cal F}
N_{e,-\sigma}(\frac{1-n_{h,\sigma}}{\epsilon_e+\epsilon_h +U_e
-2U_{eh} +i\frac{\Gamma}{2}} \nonumber
\\ +\frac{n_{h,\sigma}}{\epsilon_e+\epsilon_h +U_e+U_h
-3U_{eh}+i\frac{\Gamma}{2}})
\end{eqnarray}

and

\begin{eqnarray}
{\cal P}_2(\omega)=-\lambda {\cal F}
n_{h,\sigma}(\frac{1-N_{e,-\sigma}}{\epsilon_e+\epsilon_h +U_h
-2U_{eh}+i\frac{\Gamma}{2}} \nonumber \\
+\frac{N_{e,-\sigma}}{\epsilon_e+\epsilon_h +U_e+U_h
-3U_{eh}+i\frac{\Gamma}{2}}).
\end{eqnarray}
The first term in the filling factor ${\cal F}$ [Eq. (17)] arises
from the stimulated process, which vanishes in the vacuum state.
The second term, $N_{e,\sigma} n_{h,-\sigma}$ is due to the
spontaneous emission process. This is a quantum effect of the
electromagnetic field.  Expressions of Eqs.~(18) and (19) are
derived, respectively, from the equation of motions for $-i\langle
a^{\dagger}(t)N_{e,-\sigma}(t)d_{h,-\sigma}(t)d_{e,\sigma}(t)\rangle
$ and $-i\langle
a^{\dagger}(t)n_{h,\sigma}(t)d_{h,-\sigma}(t)d_{e,\sigma}(t)\rangle
$, which are coupled with $-i\langle
a^{\dagger}(t)n_{h,\sigma}(t)N_{e,-\sigma}(t)d_{h,-\sigma}(t)
d_{e,\sigma}(t)\rangle $. The term $-i\langle
a^{\dagger}(t)N_{e,-\sigma}(t)d_{e,\sigma}(t)d_{h,-\sigma}(t)\rangle
$ describes an electron-screened electron-hole recombination
process in a negative trion. Similarly, $-i\langle
a^{\dagger}(t)n_{h,\sigma}(t)d_{h,-\sigma}(t)d_{e,\sigma}(t)\rangle
$, describes a hole-screened electron-hole recombination process
in a positive trion, and $-i\langle
a^{\dagger}(t)n_{h,\sigma}(t)N_{e,-\sigma}(t)d_{e,\sigma}(t)d_{h,-\sigma}(t)
\rangle $ describes the biexciton-to-exciton transition process.
In the photon vacuum state, we obtain

\begin{small}
\begin{eqnarray}
&   &{\cal P}(\omega)/(N_{e,\sigma} n_{h,-\sigma})\\ \nonumber & =
& \lambda^2
\{\frac{(1-N_{e,-\sigma})(1-n_{h,\sigma})}{E_g-U_{eh}-\omega+i
\Gamma/2}
+ \frac{(1-n_{h,\sigma})N_{e,-\sigma} }{E_g+U_e-2U_{eh}-\omega +i \Gamma/2}\\
\nonumber
& + & \frac{(1-N_{e,-\sigma})n_{h,\sigma}}{E_g+U_h-2U_{eh}-\omega+i\Gamma/2}\\
\nonumber & +& \frac{N_{e,-\sigma}
n_{h,\sigma}}{E_g+U_h+U_e-3U_{eh}-\omega+i\Gamma/2}\}.
\end{eqnarray}
\end{small}
The spectrum of the polarization ${\cal P}(\omega)$ displays four
peaks, which are attributed to the exciton $X$, negative trion
$X^-$, positive trion $X^+$ and biexciton $X^2$ (biexciton
decaying to exciton). The corresponding peak positions occur at
$\omega=E_g-U_{eh}$, $\omega=E_g+U_e-2U_{eh}$,
$\omega=E_g+U_h-2U_{eh}$ and $\omega=E_g+U_e+U_h-3U_{eh}$, which
are significantly influenced by the particle Coulomb interactions.

\section{Energy levels and inter-particle interactions}
According to Eqs.~(12) and (20), particle Coulomb interactions
will significantly affect the tunnelling current of SET and the
emission spectrum of single photons. To illustrate this effect, we
apply our theory to a pyramid-shaped InAs/GaAs QD. First, we
calculate the inter-particle Coulomb interactions using a simple
but realistic effective-mass model. The electronic structures of
InAs/GaAs QDs have been extensively studied by several groups with
different methods.$^{29-33}$ The electron (hole)in the QD is
described by the equation

\begin{eqnarray}
&& [-\nabla \frac {\hbar^2} {2m_{e(h)}^*(\rho,z)} \nabla +
V^{e(h)}_{QD}(\rho,z)\mp eFz] \psi_{e(h)}({\bf r}) \nonumber \\ &=
& E_{e(h)} \psi_{e(h)}({\bf r}),
\end{eqnarray}
where ${m_e^*(\rho,z)}$ (a scalar) denotes the position-dependent
electron effective mass, which has $m_{eG}^* = 0.067 m_e$ for GaAs
and $m_{eI}^* = 0.04 m_e$ for the strained InAs in the QD.
${m_h^*(\rho,z)}$ denotes the position-dependent effective mass
tensor for the hole. It is a fairly good approximation to describe
${m_h^*(\rho,z)}$ in InAs/GaAs QD as a diagonal tensor with the
$x$ and $y$ components given by
${m^*_t}^{-1}=(\gamma_1+\gamma_2)/m_e$ and the $z$ component given
by ${m^*_l}^{-1}=(\gamma_1-2\gamma_2)/m_e$. $\gamma_1$ and
$\gamma_2$ are the Luttinger parameters. Their values for InAs and
GaAs are taken from Ref. 34. $V^e_{QD}(\rho,z)$
($V^h_{QD}(\rho,z)$) is approximated by a constant potential in
the InAs region with value determined by the conduction-band
(valence-band) offset and the deformation potential shift caused
by the biaxial strain in the QD. These values have been determined
by comparison with results obtained from a microscopic model
calculation$^{30}$ and we have $V^e_{QD}= -0.5 eV$ and
$V^h_{QD}=-0.32 eV$. The $eFz$ term in Eq.(21) arises from the
applied voltage, where $F$ denotes the strength of the electric
field. Using the eigenfunctions of Eq.~(21), we calculate the
inter-particle Coulomb interactions via
\begin{equation}
 U_{i,j}= \int d{\bf r}_1 \int d{\bf r}_2 \frac {e^2 [n_i({\bf r}_1)n_j({\bf r}_2)]}
{\epsilon_0 |{\bf r}_{1}-{\bf r}_{2}|},
\end{equation}
where $i(j)=e,h$. $n_i({\bf r}_1)$ denotes the charge density.
$\epsilon_0$ is the static dielectric constant of InAs. We have
ignored the image force arising from the small difference of
dielectric constant between InAs and GaAs. Although the Coulomb
energies are different in different exciton complexes, their
difference is rather small.$^{27,28}$ Therefore, only one set of
Coulomb interaction parameters has been used in this study.

For the purpose of constructing the approximate wave functions, we
place the system in a large confining cubic box with length L.
Here we adopt $L=40 nm$. The wave functions are expanded in a set
of basis functions, which are chosen as sine waves
\begin{small}
\begin{eqnarray}
\psi_{nlm}(\rho,\phi,z)&= &\frac{\sqrt{8}}{\sqrt{L^3}} \sin (k_lx)
\sin (k_m y)\sin (k_n z) ,
\end{eqnarray}
\end{small}
where $k_n = n\pi/L$,$k_m = m\pi/L$,$k_\ell = \ell \pi/L$. n, m
and $\ell$ are positive integers. The expression of the matrix
elements of the Hamiltonian of Eq.~(21) can be readily obtained.
In our calculation $n=20$, $m=10$, and $\ell=10$ are used in
solving  Eq.~(21). Fig. 2 shows the lowest three energy levels of
the pyramidal InAs/GaAs QD as a function of QD size. The ratio of
height and base length is fixed at $h/b=1/4$, while $h$ varies
from $2.5$ nm to $6.5$nm. Diagram (a) and (b) denote,
respectively, the energy levels for electrons and holes. We can
see that the energy level spacing between the ground state and the
first excited state is much larger than $k_B T= 2meV$, which will
be considered throughout this article. Therefore, the lowest
energy levels, $E_{e}$ and $E_{h}$, are adopted in the Hamiltonian
of Eq.~(1). The intralevel Coulomb interactions $U_{e}$ and
$U_{h}$ and interlevel Coulomb interaction $U_{eh}$ are calculated
using Eq.~(22). Fig. 3 shows the inter-particle interactions as
functions of QD size. The strengths of Coulomb interactions are
inversely  proportional to the QD size. However as the QD size
decreases below a threshold value (around $b=12 nm$), $U_{e}$ is
significantly reduced due to the leak out of electron density for
small QDs. These Coulomb interactions approach approximately the
same value in the large QD limit. This indicates similar degree of
localization for electron and hole in large QDs. We also note that
$U_{eh}$ is smaller than $U_{e}$ in the large QD. This is due to
the fact that in large QDs the degree of localization for the hole
becomes similar to that for electron, while the anisotropic nature
of hole wave function reduces $U_{eh}$. The repulsive Coulomb
interactions, $U_{e}$ and $U_{h}$, are the origin of Coulomb
blockade for electrons and holes, respectively. The attractive
Coulomb interaction $U_{eh}$ gives rise to the binding of the
exciton. To study the behavior of tunnelling current and
bias-dependent spontaneous emission spectrum, we consider a
particular pyramidal InAs/GaAs QD with base length $b=13$nm and
height $h=3.5nm$. The other relevant parameters for this QD are
$E_{e,0}=-0.14 eV$, $E_{h,0}=-0.125 eV$,$U_{e}=16.1meV$,
$U_{eh}=16.7 meV$ and $U_{h}= 18.5 meV$.

\section{Results}
In this section, we discuss our numerical results for the
tunnelling current and spontaneous emission spectrum. For
simplicity, we assume that the tunnelling rate $\Gamma_L =
\Gamma_R = 0.5 meV$ is bias-independent. We apply a bias voltage
$V_a$ across the source-drain and $V_g$ across the gate-drain.
With the applied voltages, the QD electron and hole energy levels
will be shifted to $E_e+\alpha eV_a - \beta eV_g$ and $E_h+\alpha
eV_a - \beta eV_g$, respectively. In our calculation, we assume
$\alpha =0.5 $ and $\beta = 0.7$. Meanwhile the chemical
potentials of the electrodes corresponding to a Fermi energy
$E_F=60 meV$ (relative to the conduction band minimum in the
leads) are assumed to be $70 meV$ below the energy level of
$E_{e}$ at zero bias. Other parameters adopted are: $\Gamma_h=0.2
meV$ and $R_{eh} = 10 \mu eV$.

\subsection{Effects of optical pumping on the tunnelling current}
    Applying Eqs.~(3), (13) and (15), we solve for the electron
occupation number $N_e=N_{e,\sigma}=N_{e,-\sigma}$ and the
tunnelling current $J$. Fig. 4 shows the calculated results for
$N_e$ and $J$ as functions of gate voltage with and without
photo-excitation at zero temperature and $V_a= 2 mV$. Solid line
and dashed line correspond to $I=0$ (no pump) and $I=0.9$ (with
pump), respectively. Here $I$ is a dimensionless quantity, defined
as $I\equiv \gamma_{h,c} N_{h,{\bf k}}/\Gamma_{h,s}$ which is
proportional to the pump power. The electron occupation number
displays several plateaus, while the tunnelling current displays
an oscillatory behavior. Both are typical behaviors due to the
Coulomb blockade. The manipulation of gate voltage can tune the
energy levels of the QD. For the no-pump case ($I=0$), carriers in
the emitter electrode are allowed to tunnel into the QD as the
gate voltage exceeds the threshold voltage, $V_{g3} (\approx 100
mV)$. When the energy level of $E_e+\alpha V_a - \beta V_g$ is
below the emitter chemical potential, but $E_e+\alpha V_a - \beta
V_g+U_e$ above the emitter chemical potential, $N_e$ (for a given
spin) reaches 0.5 and displays a plateau which is caused by the
Coulomb blockade associated with $U_e$ (the charging energy of the
electron ground state). At higher gate voltage, $E_e+\alpha V_a -
\beta V_g+U_e$ moves below the emitter chemical potential, and
$N_e$ approaches 1. When electron-hole pairs are generated due to
optical pumping in addition to the tunnelling electrons, various
exciton complexes such as the exciton, positive and negative
trions, and biexciton can be formed. These new channels allow
carriers in the emitter electrode to tunnel into the QD at lower
gate voltage. When the gate voltage is below $V_{g2}$, but above
$V_{g1}$, only the positive trion state is below the emitter
chemical potential, hence only a small plateau is seen. As the
gate voltage increases, more exciton complex levels are depressed
below the emitter chemical potential, and a series of plateaus
appear. The value of $N_{e}$ depends on which exciton complex
state is filled and on the pumping intensity.

 The tunnelling current as a function of the gate voltage
can be measured directly. We see that optical pumping leads to two
additional peaks below the threshold voltage $V_{g3}$, which is
caused by the electron tunnelling assisted by the presence of a
hole in the QD. This interesting phenomenon was observed by
Fujiwara et al. in an SET composed of one silicon QD and three
electrodes.$^{35}$ The behavior of the photo-induced tunnelling
current can be understood by analyzing the energy poles of the
retarded Green's function as given in Eq.~(12). In Figure 4, the
first peak of the dashed line is caused by a resonant tunnelling
through the energy pole at $\epsilon=E_e-2 U_{eh}$ (which
corresponds to a positive trion state) with probability
$(1-N_{e,-\sigma})(n_{h,\sigma}n_{h,-\sigma})$. The second peak is
caused by a pair of poles at $\epsilon=E_e - U_{eh}$ (the exciton
state) and $\epsilon=E_e+U_e-2U_{eh}$ (the biexciton state) with
probabilities $(1-N_{e,-\sigma})(n_{h,\sigma}+n_{h,-\sigma}-
2n_{h,\sigma} n_{h,-\sigma})$ and $N_{e,-\sigma}(n_{h,\sigma}
n_{h,-\sigma})$. Since the magnitude of $U_e$ is very close to
that of $U_{eh}$ for the present case, the corresponding two peaks
merge into one. For other cases (e.g. smaller QDs), the difference
between $U_e$ and $U_{eh}$ may be large enough for the peaks to be
distinguished.  The third peak is caused by another pair of poles
at $\epsilon=E_e$ (the single-electron state) and
$\epsilon=E_e+U_e -U_{eh}$ (the negative trion state) with
probabilities
$(1-N_{e,\sigma})(1-(n_{h,\sigma}+n_{h,-\sigma})+n_{h,\sigma}
n_{h,-\sigma})$ and $N_{e,-\sigma}(n_{h,\sigma}+n_{h,-\sigma}-
2n_{h,\sigma} n_{h,-\sigma})$. The last peak locating near
$V_{g}=123 mV$ is due to the resonant tunnelling through the
energy level at the pole $\epsilon=E_e+U_e$ (the two-electron
state) with probability
$N_{e,-\sigma}(1-(n_{h,\sigma}+n_{h,-\sigma})+n_{h,\sigma}n_{h,-\sigma})$.
Obviously, the relative strengths of these tunnelling current
peaks are influenced by the hole occupation numbers. The gate
voltage differences $\Delta V_{g21}=V_{g2}-V_{g1}$ and
$V_{g43}=V_{g4}-V_{g3}$ allow the determination of the
electron-hole interaction and the electron-electron Coulomb
repulsion, since $\beta \Delta V_{g21} =U_{eh}$ and $\beta
V_{g43}=U_e$. Consequently, such a measurement can be used to
determine the exciton binding energy, which is otherwise not
possible via typical photoluminescence or the electrically driven
emission spectrum measurement. In such a measurement, it is
necessary to distinguish $U_{eh}$ from $U_e$. Thus, we recommend
using a smaller bias ($V_a$), which would make the tunnelling
peaks sharper and the double-peak features at $V_g=V_{g2}$ and
$V_g=V_{g3}$ better resolved.

Figure 5 shows the tunnelling current as a function of gate
voltage for various strengths of photo-excitation power at zero
temperature.According to the definition of
$\gamma_{e(h),c}=-\sum_{\bf q}|g_{e(h),{\bf q}}|^2 Im
~1/(\gamma_{e(h),s} (\Omega_{\bf q}+i \gamma_{e(h),s}))$, the
condition $\Gamma_L=\Gamma_R \gg \gamma_{e,c}$ is still satisfied,
even though we tune $I=\gamma_{h,c} N_{h,{\bf k}}/\Gamma_{h,s}$ up
to 10 in Fig. 5. $I=10$ can be regarded as increasing $p_{exc}$ by
10 times, but it is still in the weak pumping regime. We notice
that increasing photo-excitation power tends to suppress the
tunnelling current at high gate-voltage. For instance, the peak
located at $V_{g}=123 mV$ almost vanishes due to the much reduced
probability for the pole at $\epsilon=E_e+U_e$. Based on the
results of Figs.~4 and 5, we see that the SET has potential
application as an optically controlled switch. For practical
applications, we must consider the behavior of the tunnelling
current for an SET operated at finite temperatures. Figure 6 shows
a comparison of the electron occupation number and tunnelling
current at zero temperature and finite temperature ($k_B T=2
meV$). We see that the plateaus become broadened and the magnitude
of the tunnelling current peak is reduced as the temperature
increases. Thus, a large charging energy is required for SETs to
be operated at high temperatures. The tunnelling current as a
function of applied bias ($V_a$) for various strengths of
photo-excitation power at $k_B T =2 meV$ and $V_g=0$ is shown in
Fig. 7. The tunnelling current exhibits a staircase behavior. This
is the well-known Coulomb blockade effect. The effect of the
optical pumping is reduce the threshold voltage and increase the
number of plateaus in the J-V characteristics. We also notice that
a negative differential conductance occurs at high voltages when
the resonance level of the QD is below the conduction band minimum
of the emitter electrode.

\subsection{Effects of electrodes on the spontaneous emission spectrum}
    SET has been considered as a single-photon generater (SPG) for
quantum information application such as quantum cryptography and
teleportation.$^{36}$ Thus, it is of great interest to study the
spontaneous single-photon emission spectrum of the SET. We
previously reported the theoretical studies of the emission
spectrum of a single QD embedded in a p-n junction.$^{27,28}$
Here, we consider the single-photon emission of an SET (a QD
embedded in a n-i-n junction) under optical pumping. Since the
magnitudes of inter-particle Coulomb interactions $U_e$, $U_h$ and
$U_{eh}$ are fairly close, it will be difficult to resolve the
peaks of exciton ( $X$), negative trion ($X^{-}$), positive trion
($X^{+}$) and biexciton ($X^2$) for the case of large tunnelling
rate as we considered above ($\Gamma_L=\Gamma_R=0.5 meV$).
Therefore, for this study we adopt a small tunnelling rate,
$\Gamma_L = \Gamma_R=0.1 meV$ in order to resolve these peaks in
the emission spectrum. This can be achieved, for example, by
increasing the barrier thickness between the QD and the conducting
leads. Figure 8 shows the emission spectrum for various
gate-voltages at fixed pump power ($I=0.9$). For $V_g=100mV$, the
calculated spontaneous emission spectrum displays four peaks
corresponding to $X^{-}$, $X$, $X^2$ and $X^{+}$. The red shift of
the $X^{-}$ peak relative to the exciton peak agrees well with
experimental observations.$^{16,22,24,37}$ The biexciton peak
displaying a blue shift with respect to the exciton peak (showing
an antibinding biexciton) is also consistent with the observation
reported in ref.[14,19-21] Recently, studies of the binding and
antibinding of biexcitons were reported in Ref. 21. Our
calculation given by Eqs.~(21) and (22) provides only the
antibinding feature of biexciton. In Ref. 21 it is pointed out
that the biexciton complex changes from antibinding to binding as
the  QD size increases. For QDs with dimension larger than the
exciton Bohr radius (around 20nm), the correlation energy becomes
significant. In this study, we have not taken into account the
correlation energy. This is justified as long as we restrict
ourselves to QDs with size less than 20nm. We see in this figure
when the gate-voltage is increased up to $V_g=125 mV$, only the
$X^{-}$ and $X^{2}$ peaks survive in the emission spectrum as
$N_e$ approaches 1. Furthermore, we can adjust the intensity of
excitation power to either enhance or suppress the strength of the
biexciton peak. Thompson et al. demonstrated that the single
photon generated by the biexciton state is significantly less in
emission time than the single exciton state.$^{19}$ From this
point of view, the biexciton state is favored for the application
of single-photon generation. We find that one can increase the
pump power in order to select the biexciton state as the desired
source to produce the single-photon emission. This is demonstrated
in Fig.~9, where the emission spectra for various strengths of
excitation power at the fixed gate voltage $V_g=125 mV$ are shown.
We see that the negative trion saturates at $I=1.9$ and diminishes
afterwards, while the biexciton peak increases quadratically with
$I$ and becomes the dominant peak in the emission spectrum.

Finally, we discuss the emission spectrum of an isolated
charge-neutral QD with $\Gamma_L =\Gamma_R =0$. For this case the
rate equation of exciton number $N_{X,\sigma}$ can be readily
obtained by solving the rate equation
\begin{equation}
\gamma_{X,c} N_k (1-N_{X,\sigma})- \Omega \int \frac{d
\omega}{\pi} \omega \rho(\omega) |ImP(\omega)|=0
\end{equation}
with
\begin{equation}
P(\omega)=
N_{X,\sigma}(\frac{1-N_{X,-\sigma}}{E_X-\omega+i\Gamma/2}
+\frac{N_{X,-\sigma}}{E_X+\Delta E_c -\omega+i\Gamma/2}),
\end{equation}
where $\Omega=4\pi n^3_r \mu^2_r/(6c^3\hbar^3\epsilon_0)$ and
$n_r$ is the refractive index of the system.
$\rho(\omega)=\omega^2$ arises from the density of states of
photons. $E_X$ and $\Delta E_{X,c}$ denote, respectively, the
exciton energy and exciton-exciton correlation energy.
$\gamma_{X,c}$ denotes the capture rate for excitons from the
wetting layer to the QD. After the integration of $\omega$ in  Eq.
(24), we obtain
\begin{equation}
\Omega N_{X,\sigma} ( (1-N_{X,-\sigma})E^3_X +N_{X,-\sigma}
(E_X+\Delta E_c)^3)
\end{equation}
for small $\Gamma$, and
\begin{equation}
N_{X,\uparrow}=N_{X,\downarrow}=\frac{\gamma_{X,c}N_k}{\gamma_{X,c}N_k+\gamma_{X,r}},
\end{equation}
where $\gamma_{X,r}=\Omega E^3_X$ denotes the exciton decaying
rate, and we have ignored $\Delta E_c$, since it is small compared
to $E_X$.  According to Eq.~(25), we see that the intensities of
exciton and biexciton peaks are proportional to $p_{exc}$ and
$p^2_{exc}$, if $N_{X,\sigma}$ is proportional to $p_{exc}$ (the
excitation power). This is in very good agreement with many
experimental reports.$^{17,19-21,38}$ Note that in some
experiments, the pump energy is far above the QD band gap and the
emission occurs after the electrons and holes are captured. This
situation becomes similar to the open system considered in the SET
with optical pumping. In this case, we can replace $N_{e,\sigma}$
in Eq.~(20) by $n_{e,\sigma}$, the electron occupation due to
optical pumping, and we can still prove that the intensities of
$X$ and $X^2$ peaks as functions of the excitation power are
described by Eqs.~(14), (15) and (20). Thus, they exhibit linear
and quadratic behavior as a function of $p_{exc}$, respectively.
So, there is no qualitative difference between Eq.~(20) and
Eq.~(24) regarding the behavior of $X$ and $X^2$, except that
Eq.~(20) also describes the possibility of emission due to
recombination in trions. This provides an explanation to the
experimental observation of Ref. 17, in which the pump energy is
far above the QD band gap and both $X^{-}$ and $X^{+}$ peaks (in
addition to the $X$ and $X^2$ peaks) were observed in the emission
spectrum.

\section{Summary}
In this article we have studied the tunnelling current and
emission spectrum of an SET under optical pumping theoretically.
We apply our theoretical analysis to an SET, which consists of a
single quantum dot embedded in a n-i-n junction. We use a simple
but realistic effective mass model to calculate the inter-particle
Coulomb interactions of pyramid-shaped InAs/GaAs QDs. These
Coulomb interactions are essential for the study of optical and
transport properties of QDs. The retarded Green's function was
calculated via a non-equilibrium Green's function method. It is
found that the holes in QD generated by optical pumping leads to
new channels for electrons to tunnel from the emitter to the
collector, which creates a variety of ways to control both the
electrical and optical signals. The binding energy of exciton
complexes as well as electron charging energy can be determined by
examining the tunnelling current as a function of gate voltage
$V_g$ with a small applied bias $V_a$. In addition, the emission
spectrum of SET can be modified significantly by adjusting bias
voltage, the gate voltage, or the pump intensity.

{\bf ACKNOWLEDGMENTS}

This work was supported by National Science Council of Republic of
China under Contract Nos. NSC 93-2215-E-008-014,
NSC-93-2120-M-008-002, and NSC 93-2215-E-008-011.

\section{Appendix A}
In the system described by Eq. (1), electromagnetic field of
frequency $\omega_0$ is used to resonantly pump carriers into the
InAs wetting layer. Therefore, electrons can be injected into the
InAs QD via either the tunnelling process or carrier capture from
the wetting layer. On the other hand, holes in InAs QD are
provided only by the capture process. The dominant carrier capture
process is the the optical-phonon assisted carrier relaxation as
described by the fifth term in Eq. (1). Because of the
phonon-bottleneck effect, the carrier capture rates for QDs are
suppressed. Consequently, the condition $\Gamma_e \gg
\gamma_{e,c}$ is usually satisfied, where $\Gamma_e$ and
$\gamma_{e,c}$ denote the electron tunnelling rate and electron
capture rate, respectively. Based on the above condition, we can
ignore the electron capture process in the calculation of the
tunnelling current. The tunnelling current from the left electrode
to the InAs QD can be calculated from the time evolution of the
occupation number, $N_{L}=\sum_{{\bf k}, \sigma;L}
c^{\dagger}_{{\bf k},\sigma;L} c_{{\bf k},\sigma;L}$.

Before the calculation of tunnelling current, an unitary
transformation is employed to remove the phase of the optical
pumping term. The renormalized energy levels of the electron and
hole become $\epsilon_e=E_e-\omega/2$ and
$\epsilon_h=E_h-\omega/2$. Furthermore, the hopping terms between
the leads and the dot become time dependent, $V_{{\bf
k},j}(t)=V_{\bf k} exp[-(\pm i\omega t)/2]$. Now, we can apply the
formulas given in ref.[5]. We obtain

\begin{equation}
J=\frac{-2e}{\hbar} \int \frac{d\epsilon}{2\pi} \Gamma_L(\epsilon)
[\frac{1}{2}G^{<}_{e,\sigma}(\epsilon)+f_L(\epsilon)ImG^{r}_{e,\sigma}(\epsilon)],
\end{equation}
where $\Gamma_L(\epsilon)=2\pi\sum_{\bf k} |V_{L,{\bf k}}|^2
\delta(\epsilon-\epsilon_{\bf k})$, and $f_L(\epsilon)$ is the
Fermi distribution function. $G^{<}_{e,\sigma}(\epsilon)$ and
$G^{r}_{e,\sigma}(\epsilon)$ are the Fourier transformation of the
lesser Green function $G^{<}_{e,\sigma}(t)=i\langle
d^{\dagger}_{e,\sigma}(0) d_{e,\sigma}(t)\rangle$ and retarded
Green's function
$G^{r}_{e,\sigma}(t)=-i\theta(t)\langle\{d_{e,\sigma}(t),d^{\dagger}_{e,\sigma}(0)\}\rangle$.
Applying Dyson's  equation, we obtain
\begin{equation}
G^{<}_{e,\sigma}(\epsilon)=
f^{<}(\epsilon)[G^{r}_{e,\sigma}(\epsilon)-G^{a}_{e,\sigma}(\epsilon)]
\end{equation}
with
\begin{equation}
f^{<}(\epsilon)=\frac{i \sum_{{\bf k},\ell}|V_{{\bf k},\ell}|^2
g^{<}_{e,{\bf k},\sigma;\ell}(\epsilon)+i \sum_Q |\lambda|^2 {\cal
G}^{<}_{h,-\sigma}(\epsilon)}{-i \sum_{{\bf k},\ell}|V_{{\bf
k},\ell}|^2 (g^{r}_{e,{\bf k},\sigma;\ell}(\epsilon)-g^{a}_{e,{\bf
k},\sigma;\ell}(\epsilon))-i \sum_Q |\lambda|^2 Im{\cal
G}^{r}_{h,-\sigma}(\epsilon)},
\end{equation}
where $g^{<}_{e,{\bf k},\sigma;\ell}(\epsilon)=f_{\ell}(\epsilon)
(g^{r}_{e,{\bf k},\sigma;\ell}(\epsilon)-g^{a}_{e,{\bf
k},\sigma;\ell}(\epsilon))$.$(g^r_{\ell}(\epsilon)-g^{a}_{\ell}(\epsilon))=i2\pi
\delta(\epsilon-\epsilon_{\bf k})$. The lesser Green's function
and retarded Green's function for holes are denoted by ${\cal
G}^{<}_{h,-\sigma}(\epsilon)$ and ${\cal
G}^{r}_{h,-\sigma}(\epsilon)$. Due to the small electron-hole
recombination rate $R_{eh} \approx 1/ns$, terms involving
$|\lambda|^2$ can be droped. That is to ignore the photocurrent in
Eq. (28). After rewriting $f^{<}(\epsilon)$ as
\begin{equation}
f^{<}(\epsilon)=-\frac{\Gamma_L(\epsilon)f_L(\epsilon)+\Gamma_R(\epsilon)
f_R(\epsilon)}{\Gamma_L(\epsilon)+\Gamma_R(\epsilon)},
\end{equation}
and substituting it into Eq. (28), we obtain the tunnelling
current
\begin{equation}
J=\frac{-2e}{\hbar} \int \frac{d\epsilon}{2\pi}
\frac{\Gamma_L(\epsilon)
\Gamma_R(\epsilon)}{\Gamma_L(\epsilon)+\Gamma_R(\epsilon)}[f_L(\epsilon)-f_R(\epsilon)]
ImG^{r}_{e,\sigma}(\epsilon).
\end{equation}

The expression of Eq. (32) was also obtained by Meir, Wingreen and
Lee$^{6}$. It is of no surprise that we reproduce their result,
because the system considered here reduces to their case as the
carrier relaxation process and electron-hole recombination process
are both turned off. Note that the retarded Green's function
$G^{r}_{e,\sigma}(\epsilon)$ includes the full effects due to the
electron-electron Coulomb interaction and electron-hole Coulomb
interaction and the effect due to the dot-electrodes coupling to
the lowest order. Also note that the electrodes are coupled to the
conduction band of the QD, but not to the valence band of the QD.

\section{Appendix B}
When light is used to resonantly create electrons and holes in the
wetting layer, electrons and holes can relax to the InAs QD via
optical phonon-assisted process. When electrons relax into InAs
QD, they will tunnel out quickly since the electrons in the ground
state of the InAs QD are coupled with the electrodes. However,
holes in the InAs QD can decay only via recombination with
electrons (radiation process) or via impurity scattering
(nonradiation process). When holes are present in the InAs QD, we
expect the tunnelling current to change significantly due to the
large electron-hole Coulomb interactions [see Eq. (12)]. In this
appendix, we give the derivations for Eqs. (14) and (15), which
describe the electron and hole populations in the InAs QD under
optical pumping. First, we introduce the electron and hole
distribution functions in the wetting layer, $N_{e,{\bf
k}}=\langle d^{\dagger}_{e,{\bf k}}(t) d_{e,{\bf k}}(t)\rangle$
and $N_{h,-{\bf k}}=\langle d^{\dagger}_{h,-{\bf k}}(t) d_{h,-{\bf
k}}(t)\rangle$. We use the equation-of-motion method to calculate
$N_{e,{\bf k}}=\langle d^{\dagger}_{e,{\bf k}}(t) d_{e,{\bf
k}}(t)\rangle$ and $N_{h,-{\bf k}}=\langle d^{\dagger}_{h,-{\bf
k}}(t) d_{h,-{\bf k}}(t)\rangle$. We obtain
\[
\frac{d}{dt}N_{e,{\bf k}}=-i\gamma_{e,s} N_{e,{\bf k}}+ 2Im\langle
\lambda^{\dagger}_0 b^{\dagger}_{e,{\bf k}} b^{\dagger}_{h,-{\bf
k}}\rangle \]
\begin{equation}
-2\sum_{\bf q} Im ~g_{e,{\bf q}} \langle A b^{\dagger}_{e,{\bf k}}
d_{e,\sigma}\rangle,
\end{equation}
and
\[
\frac{d}{dt}N_{h,-{\bf k}}=-i\gamma_{h,s} N_{h,-{\bf k}}+
2Im\langle \lambda^{\dagger}_0 b^{\dagger}_{e,{\bf k}}
b^{\dagger}_{h,-{\bf k}}\rangle \]
\begin{equation}
-2\sum_{\bf q}Im g_{h,{\bf q}}
\langle A b^{\dagger}_{h,-{\bf k}} d_{h,-\sigma}\rangle,
\end{equation}
where $\gamma_{e(h),s}$ denotes the carrier scattering rate due to
mechanisms not considered in the Hamiltonian. $\langle
\lambda^{\dagger}_0 b^{\dagger}_{e,{\bf k}} b^{\dagger}_{h,-{\bf
k}}\rangle$  represents the polarization for creating an
electron-hole pair in the wetting layer. $g_{e,{\bf q}} \langle A
b^{\dagger}_{e,{\bf k}} d_{e,\sigma}\rangle $ ($g_{h,{\bf q}}
\langle A b^{\dagger}_{h,-{\bf k}} d_{h,-\sigma}\rangle$) are
terms due to the process for transferring one electron (hole) from
the QD to the wetting layer by absorbing one optical phonon. To
terminate the equations of motion [Eqs. (33) and (34)], we need to
solve $\langle \lambda^{\dagger}_0 b^{\dagger}_{e,{\bf k}}
b^{\dagger}_{h,-{\bf k}}\rangle$ and $g_{e,{\bf q}} \langle A
b^{\dagger}_{e,{\bf k}} d_{e,\sigma}\rangle $ ($g_{h,{\bf q}}
\langle A b^{\dagger}_{h,-{\bf k}} d_{h,-\sigma}\rangle$). The
equation of motion for $g_{e,{\bf q}} \langle A
b^{\dagger}_{e,{\bf k}} d_{e,\sigma}\rangle$ leads to
\[
\frac{d}{dt}g_{e,{\bf q}} \langle A b^{\dagger}_{e,{\bf k}}
d_{e,\sigma}\rangle=(i\Omega_q-\gamma_{e,s})\langle A
b^{\dagger}_{e,{\bf k}} d_{e,\sigma}\rangle +|g_{e,{\bf q}}|^2 \]
\begin{equation}
[(1+n_q) (1-(n_e+N_e))N_{e,{\bf k}} -n_q (n_e+N_e)(1-N_{e,{\bf
k}})],
\end{equation}
where $n_q$ denotes the phonon distribution function, defined as
$n_q(\omega_q)=1/(exp(\omega_q/k_bT)-1)$. $n_e=\langle
d^{\dagger}_{e,\sigma}(t) d_{e,\sigma}(t)\rangle_p $ and
$N_e=\langle d^{\dagger}_{e,\sigma}(t) d_{e,\sigma}(t)\rangle_t$
denote the electron occupation number of the QD due to the pumping
process and tunnelling process, respectively.
$\Omega_q=(\epsilon_{\bf k}-E_e-\omega_q)$. Meanwhile, the
polarization obeys the following equation
\[
\frac{d}{dt}\langle \lambda^{\dagger}_0 b^{\dagger}_{e,{\bf k}}
b^{\dagger}_{h,-{\bf k}}\rangle
=(-i\Omega_0+\gamma_{e,s}+\gamma_{h,s})\langle \lambda^{\dagger}_0
b^{\dagger}_{e,{\bf k}} b^{\dagger}_{h,-{\bf k}}\rangle \]
\begin{equation}
-i |\lambda_0|^2 (1-N_{e,{\bf k}}-N_{h,-{\bf k}}),
\end{equation}
where $\Omega_0=\epsilon_{e,{\bf k}}+\epsilon_{h,-{\bf
k}}-\omega_0$. In the steady state, we obtain
\[
N_{e,{\bf k}}=p_{exc} (1-N_{e,{\bf k}}-N_{h,-{\bf
k}})-\gamma_{e,c} [(1+n_q) (1-(n_e+N_e)) \]
\begin{equation}
N_{e,{\bf k}}-n_q
(n_e+N_e) (1-N_{e,{\bf k}})],
\end{equation}
where $p_{exc}\equiv
-Im|\lambda_0|^2/(\gamma_{e,s}(\Omega_0+i(\gamma_{e,s}+\gamma_{h,s})))$
and $\gamma_{e,c}=-\sum_{\bf q}Im|g_{e,{\bf q}}|^2/(\gamma_{e,s}
(\Omega_q+i\gamma_{e,s}))$. For low pumping intensity, $N_{e,{\bf
k}}$ is linearly proportional to $p_{exc}$. In addition, we note
that $\gamma_{e,c}$ is fairly small due to the phonon bottleneck
effect of the QD. Similarly for holes in the wetting layer, we
obtain
\[
N_{h,-{\bf k}}=p_{exc} (1-N_{e,{\bf k}}-N_{h,-{\bf
k}})-\gamma_{h,c} [(1+n_q) (1-n_h) \]
\begin{equation}
N_{h,-{\bf k}}-n_q n_h
(1-N_{h,-{\bf k}})],
\end{equation}
where $\gamma_{h,c}=-\sum_{\bf q}Im|g_{h,{\bf q}}|^2/(\gamma_{h,s}
(\Omega_q+i\gamma_{h,s}))$ is the hole capture rate. Next, we need
to solve the equation of motion for $n_{e,\sigma}$ and
$n_{h,-\sigma}$. We have
\[
\frac{d}{dt}n_{e,\sigma}=-i\Gamma_{e} n_{e,\sigma}+ \sum_{\bf Q}
2Im\langle \lambda^{\dagger} a d^{\dagger}_{e,\sigma}
d^{\dagger}_{h,-\sigma}\rangle \]
\begin{equation}
+2Im g_{e,{\bf q}} \langle A b^{\dagger}_{e,{\bf k}}
d_{e,\sigma}\rangle,
\end{equation}
and
\[
\frac{d}{dt}n_{h,-\sigma}=-i\Gamma_{h,s} n_{h,-\sigma}+ \sum_{\bf
Q} 2Im\langle \lambda^{\dagger} d^{\dagger}_{e,\sigma}
d^{\dagger}_{h,-\sigma}\rangle \]
\begin{equation}
+2Im g_{h,{\bf q}} \langle A
b^{\dagger}_{h,-{\bf k}} d_{h,-\sigma}\rangle,
\end{equation}
where $\Gamma_e=\Gamma_L+\Gamma_R$ and $\Gamma_{h,s}$ denote the
electron tunnelling rate and  hole-impurity scattering rate,
respectively. At low temperatures, the phonon-absorption process
is negligible, since $n_q \rightarrow 0$. Using Eqs. (35) and
(20), we obtain the steady-state solution for $n_{e,\sigma}$
\begin{equation}
n_{e,\sigma}=\frac{\gamma_{e,c}N_{e,{\bf
k}}(1-N_e)}{\gamma_{e,c}N_{e,{\bf k}}+R_{eh}n_h+\Gamma_e}.
\end{equation}
Note that the electron-hole recombination rate $R_{eh}$ is given
by
\begin{equation}
R_{eh}= \Omega \int \frac{d\omega}{\pi} \rho(\omega) |Im{\cal
P}(\omega)|,
\end{equation}
where $\Omega=4\pi n^3_r \mu^2_r/(6c^3\hbar^3\epsilon_0)$ and
$n_r$ is the refractive index of the system.
$\rho(\omega)=\omega^2$ arises from the density of states of
photons. Similarly, we obtain
\begin{equation}
n_{h,-\sigma}=\frac{\gamma_{h,c}N_{h,{\bf
k}}}{\gamma_{h,c}N_{h,-{\bf k}}+R_{eh}(n_e+N_{e})+\Gamma_{h,s}}.
\end{equation}
Because of the condition $\Gamma_e \gg \gamma_{e,c}$ and weak
pumping intensity, we have $N_{i,{\bf k}}\approx p_{exc}$, and
$n_{e}$ is negligible. As for $N_e$, we can employ Eqs. (29) and
(31) to obtain
\begin{equation}
N_{e,\sigma}=-\int \frac{d\epsilon}{\pi} \frac{\Gamma_L
f_L(\epsilon)+\Gamma_R f_R(\epsilon)}{\Gamma_L+\Gamma_R}
ImG^r_{e,\sigma}(\epsilon).
\end{equation}

\mbox{}

\newpage

{\bf Figure Captions}

Fig. 1: Schematic diagram of a single electron transistor under
optical pumping. $\Gamma_L $ and $\Gamma_R$ denote, respectively,
the tunnelling rate for electrons in the QD to the source and
drain electrodes. $R_{eh}$ denotes the electro-hole recombination
rate. The vertical wavy line indicates the optical pumping, and
the horizontal wavy line indicates the single-photon emission.

Fig. 2: The lowest three energy levels of a quantum dot as
functions of the QD size $b$ for (a) electrons and (b) holes.

Fig. 3: Intralevel Coulomb interactions $U_{e}$ and $U_{h}$ and
interlevel Coulomb interaction $U_{eh}$ as functions of the QD
base length $b$.

Fig. 4: Electron occupation number $N_e$ and tunnelling current as
functions of gate voltage at zero temperature for various
strengths of optical excitation. Current density is in units of
$J_0=2e \times meV/h$.

Fig. 5: Tunnelling current as a function of gate voltage at zero
temperature for various strengths of optical excitation.

Fig. 6: Electron occupation number $N_e$ and tunnelling current as
functions of gate voltage for various temperatures at fixed
optical excitation strength ($I=0.9$). Current density is in units
of $J_0=2e \times meV/h$.

Fig. 7: Tunnelling current as a function of applied bias at
temperature $k_B T = 2 meV$ for various strengths of optical
excitation. Current density is in units of $J_0=2e \times meV/h$.

Fig. 8: Intensities of emission spectrum for various gate voltages
at fixed optical excitation strength ($I=0.9$).

Fig. 9: Intensities of emission spectrum for various strengths of
optical excitation at fixed gate voltage ($V_g=125 mV$).


\begin{thebibliography}{50}

\bibitem[1]{Gor} D. Goldhabar-Gordon. H. Shtrikman, D. Mahalu, D. Abusch-magder,
U. Meirav, and M. A. Kastner, Nature {\bf 391}, 156 (1998).

\bibitem[2]{Gro} S. M. Cronenwett, T. H. Oosterkamp, and L. P. Kouwenhoven,
Science \textbf{281},  540 (1998).

\bibitem[3]{Sco}J. H. F. Scott-Thomas, S. B.
Field, M. A. Kastner, H. I. Smith, and D. A. Antoniadis, Phys.
Rev. Lett, \textbf{62},  583 (1989).

\bibitem[4]{Mei} U. Meirav, M. A. Kastner, and S. J. Wind, Phys. Rev. Lett.
\textbf{65}, 771 (1990).

\bibitem[5]{Jau} A. P. Jauho, N. S. Wingreen, and Y. Meir, Phys. Rev. B \textbf{50},
5528 (1994).

\bibitem[6]{Mei1}Y. Meir, N. S. Wingreen, and P. A. Lee, Phys. Rev. Lett.\textbf{70},
2601 (1993).


\bibitem[7]{Hag} H. Haug and A. P. Jauho, Quantum Kinetics in Transport and
Optics o Semiconductor (Springer, Heidelberg, 1996).

\bibitem[8]{Zhu} L. Zhuang, L. Guo, and S. Y. Chou, Appl. Phys. Lett. \textbf{72},
1205 (1998).

\bibitem[9]{Sai} M. Saitoh, N. Takahashi, H. Ishikuro and T. Hiramoto, Jpn. J.
Appl. Phys.\textbf{ 40}, 2010 (2001).

\bibitem[10]{Tan} Y. T. Tan, T. Kamiya, D. Zak, H. Ahmed, J. Appl. Phys. \textbf{94}, 633
(2003).

\bibitem[11]{Li} P. W. Li, W. M. Liao, D. M. T. Kuo, S. W. Lin, P. S. Chen,
S. C. Lu, and M. J. Tsai, Appl. Phys. Lett., 85, 1532 (2004).

\bibitem[12]{Tak} T. Takagahara and K. Takeda, Phys. Rev. B \textbf{46},
15578 (1992).

\bibitem[13]{Mic}P. Michler, A. Kiraz, C. Becher, W. V. Schoenfeld, P. M. Petroff,
L. D. Zhang, E. Hu and A. Imamoglu: Science \textbf{290} 2282
(2000).

\bibitem[14]{Yua} Z. Yuan, B. E. Kardynal, R. M. Stevenson, A. J. Shield, C. J.
Lobo, K. Cooper, N. S. Beattlie, D. A. Ritchie and M. Pepper:
Science \textbf{295}, 102 (2002).

\bibitem[15]{Ima} A. Imamoglu, and Y. Yamamoto: Phys. Rev. Lett.
{\bf 72},  210 (1994).

\bibitem[16]{Lan} L. Landin, M. S. Miller, M. E. Pistol, C. E.
Pryor, and L. Samuelson, Science, \textbf{280}, 262 (1998).

\bibitem[17]{San} C. Santori, M. Pelton, G. Solomon, Y. Dale, and
Y. Yamamoto, Phys. Rev. Lett. \textbf{86}, 1502 (2001).

\bibitem[18]{San1} C. Santori, D. Fattal, M. Pelton, G. S. Solomon,
and Y. Yamamoto, Phys. Rev. B \textbf{66}, 045308 (2002).

\bibitem[19]{Tho} R. M. Thompson, R. M. Stvenson, A. J. Shields,
I. Farrer, C. J. Lobo, D. A. Ritchie, M. L. Leadbeater, and M.
Pepper, Phys. Rev. B \textbf{64}, 201302 (2001).


\bibitem[20]{Mor} E. Moreau, I. Robert, L. Manin, V. Thierry-Mieg,
J. M. Gerard, and I. Abram, Phys. Rev. Lett. \textbf{87} 18360,
(2001).
\bibitem[21]{Rod} S. Rodt, R. Heitz, A. Schliwa, R. L. Sellin, F.
Guffarth, and D. Bimberg, Phys. Rev. B \textbf{68}, 035331 (2003).

\bibitem[22]{War} R. J. Warburton, C. Schaflein, D. Haff, F.
Bickel, A. Lorke, K. Karral, J. M. Garcla, W. Schoenfeld, and P.
M. Petroff, Nature, \textbf{405} 926 (2000).

\bibitem[23]{Hog} A. Hogele, S. Seidl, M. Kroner, K. Karrai, R. J.
warburton, B. D. Gerardot, and P. M. Petroff, Phys. Rev. Lett.
\textbf{93}, 21704 (2004).

\bibitem[24]{Fin} F. Findeis, M. Baier, E. Beham, A. Zrenner, and
G. Abstreiter, Appl. Phys. Lett.  \textbf{78}, 2958 (2001).

\bibitem[25]{Cha} Y. C. Chang and David. M. T. Kuo, Appl. Rev.
Lett,\textbf{ 83}, 156 (2003).

\bibitem[26]{Kuo} David. M. T. Kuo, Jap. J. Appl. Phys. \textbf{43}, L 1607
(2004).

\bibitem[27]{Kuo1} David. M. T. Kuo and Y. C. Chang, Phys. Rev. B
{\bf  69}, 041306 (2004).

\bibitem[28]{Kuo2} David. M. T. Kuo and Y. C. Chang, J. Phys. Soc.
Jap. \textbf{74} 690 (2005).

\bibitem[29]{Kel} L. V. Keldysh, Zh. Eksp, Teor. Fiz. {\bf 47},
1515 (1964)[Sov. Phys. JETP {\bf 20},  1018 (1965)].

\bibitem[30]{Sti} O. Stier, M. Grundmann, and D. Bimberg, Phys.
Rev. B \textbf{59}, 5688 (1999).

\bibitem[31]{Sun} S. J. Sun, and Y. C. Chang: Phys. Rev. B
\textbf{62}, 13631 (2000).

\bibitem[32]{Pry} C. Pryor, Phys. Rev. B \textbf{57}, 7190 (1998).

\bibitem[33]{wan} L. W. Wang, J. Kim and A. Zunger, Phys. Rev. B
\textbf{59}, 5678 (1999).

\bibitem[34]{Law} P. Lawaetz, Phys. Rev. B \textbf{4}, 3460 (1971).


\bibitem[35]{Fuj} A. Fujiwara, Y. Takahashi and K. Murase, Phys.
Rev. Lett. \textbf{78} 1532 (1997).

\bibitem[36]{Vah} K. J. Vahala, Nature \textbf{424}, 839 (2003).

\bibitem[37]{Har} A. Hartmann, Y. Ducommun, E. Kapon, U.
Hohenester and E. Molinari, Phys. Rev. Lett. \textbf{84}, 5648
(2000).

\bibitem[38]{Bru} K. Brunner, G. Abstreiter, G. Bohm, G. Trankle,
and G. Weimann, Phys. Rev. Lett \textbf{73}, 1138 (1994).


\end{thebibliography}
\end{document}